# Evolution of structural properties of iron oxide nano particles during temperature treatment from 250°C – 900°C: X-ray diffraction and Fe K-shell pre-edge X-ray absorption study

Debajeet K. Bora[1, 2], Artur Braun [1,3*], Selma Erat [1, 4], Olga Safonova [5],
Thomas Graule[1, 6], Edwin C. Constable [2]

[1]Laboratory for High Performance Ceramics

Empa, Swiss Federal Laboratories for Material Science and Technology

CH-8600 Dübendorf, Switzerland

[2]Department of Chemistry, University of Basel, CH-4056 Basel, Switzerland

[3]Hawaii Natural Energy Institute, School of Ocean and Earth Science and Technology

University of Hawaii at Manoa, Honolulu, HI 96822, USA

[4]Department of Nonmetallic Inorganic Materials

ETH Zürich-Swiss Federal Institute of Technology, CH-8037 Zürich, Switzerland

[5]Swiss Norwegian Beamline, European Synchrotron Radiation Facility (ESRF)

Grenoble-38043, France

[5]Technische Universität Bergakademie Freiberg

---

[*] Corresponding author, Tel: +41(0)44 823 4850, Fax: +41(0)44 823 4150, artur.braun@alumni.ethz.ch

D-09596 Freiberg, Germany

## ABSTRACT

Iron oxide nano particles with nominal $Fe_2O_3$ stoichiometry were synthesized by a wet, soft chemical method with the heat treatment temperatures from 250°C to 900°C in air. The variation in the structural properties of the nano particles with the heat treatment temperature was studied by X-ray diffraction and Fe K shell X –ray absorption study. X-ray diffractograms show that at lower annealing temperatures nano particle comprises both maghemite and hematite phases. With increasing temperature, the remainder of the maghemite phase transformed completely to hematite. Profile analysis of the leading Bragg reflections reveals that the average *crystallite size* increases from 50 nm to 150 nm with increasing temperature. The mean primary *particle size* decreased from 105 nm to 90 nm with increasing heat treatment temperature**.** The X-ray diffraction results are paralleled by systematic changes in the pre-edge structure of the Fe K-edge X-ray absorption spectra, in particular by a gradual decrease of the $t_{2g}/e_g$ peak height ratio of the two leading pre-edge resonances, confirming oxidation of the Fe from $Fe^{2+}$ towards $Fe^{3+}$. Transmission electron microscopy (TEM) on the samples treated at temperatures as high as 900°C showed particles with prismatic morphology along with the formation of stacking fault like defects. High resolution TEM with selected area electron diffraction (SAED) of samples heat treated above 350°C showed that the nano particles have well developed lattice fringes corresponding to that of (110) plane of hematite.

## I. Introduction

Nano sized iron oxide particulates have emerged as versatile materials for different applications due to their, magnetic, electronic, photonic and optical properties. The structure-function relationship of these nano particles have been intensively studied because of the applications in magnetic storage, gas sensing, biomedical, and catalysis applications [1-5]. Iron oxides nano particles have been prepared by a variety of methods such as sonochemical reactions [6], mechanochemical synthesis [7], hydrolysis, thermolysis of precursors as well as co-precipitation technique [8]. Nano particles with a virtually near to monodisperse size distribution can be produced by thermal decomposition of an iron-cupferron complex in octylamine [9]. Organometallic complexes of iron are being used as precursors for the synthesis. For example, iron pentacarbonyl, $Fe(CO)_5$, with oleic acid and trimethyl oxide in octyl ether has been used for maghemite synthesis [10]. Out of various phases of iron oxide nano particle recently great interest has been devoted towards to the synthesis of α-phase of iron oxide nano particles (α-$Fe_2O_3$, hematite). These are of technological interest for the use in photoelectrochemical (PEC) water splitting reaction for the production of hydrogen [11-18]. Hematite is the inexpensive material of interest for PEC application due to its suitable band gap (2.2 eV), valence band edge position, earth abundance and environmentally benign nature [19]. Hematite nano particles can be prepared by several techniques such as hydrothermal approach, catalytic synthesis, flame spray pyrolysis, non aqueous synthesis, surfactant assisted method such as CTAB etc. [20-25]. Besides these, a versatile and low cost technique was developed to synthesize hematite nano particles through a gentle chemistry route, which has some advantages of its own [26]. Unlike some processes where large aggregates are formed, occurrence of uncontrolled oxidation, and presence of matrix etc. the new technique is free from all of these problems. The main characteristic of this soft chemistry route is the lower temperature of the thermal treatment. This technique also ensures the presence of uniform particle size distribution. The procedure is characterized by a complete and homogeneous mixing of initial ingredients molecular or atomic level. As a result the synthesized product is characterized by their small particle size, high homogeneity as well as stoichiometry.

Considering the crystallographic structure of hematite, it is isostructural to a corundum structure wherein oxygen atoms are closely packed with the same density around the $Fe^{3+}$ cations, but with a slightly different stacking, which may therefore result in slight distortions from a perfect octahedral environment. There are six formula units per unit cell and possess hexagonal symmetry. $Fe^{III}$ ions which are arranged regularly with two filled sites filled two third of the site. In the next step, one vacant site from (001) plane followed the same and ultimately forming the six fold rings. Its structure consists of hexagonally closed packed (hcp) arrays of oxygen ions along the [001] direction whereby (001) plane is parallel with plane of anions [27] .The unique properties of hematite nano particles are  mostly associated with the oxidation state and coordination geometry, which can be probed directly from the X-ray absorption spectroscopy. The X-ray absorption near-edge structure (XANES) is sensitive to local geometries and electronic structures of atoms. Both from a theoretical and an experimental point of view, XANES gives information on a specifically local environment and hence is a powerful tool to gain insight into the short- and medium-range structure. These techniques provide information about the oxidation state, the symmetry of the local environment, the partial density of unoccupied states, the coordination number, and the distance to the next neighbour [28-31]. X-ray absorption is a local process in which an electron is promoted to an excited electronic state, which can be coupled to the original core level by the dipole selection rule ($\Delta l = \pm 1$). The pre-edge peaks in XANES spectra are mostly attributed to $1s$ to $nd$ quadrupole allowed transitions in the six fold-coordinated cation systems due to distortion [32]. The variation in the pre-edge peak absorption features both in intensity and energy position in the cation $K$-edge XANES spectra provides ample evidence for the coordination of transition metal which displays a complex behaviour in crystalline and amorphous solids, minerals, biological molecules, and cations adsorbed at surfaces.

The main objective of the study is to understand the evolution in structural properties of iron oxide nano particles formed at different thermal treatment from 250°C – 900°C. For evaluating it, we first synthesized the iron oxide nano particles by the heat treatment of iron fatty acid precursor complex prepared by using a non-aqueous chemical route [25] at different temperature from 250°C to 900°C. In the next step, nano particles have been characterized by X-ray Diffractometry and Fe K - edge XANES spectroscopy to get information about the structural evolution. During the study, our main focus will be primarily on the hematite nano particles which are considered as a photoactive material in the photoelectrochemical water splitting application. The thermal treatment during the processing of the hematite film has a profound influence on the final photoelectrochemical functionality of the material due to the change in structure of the nano particle in the film [33]. Also, the mixing with other phases such as maghemite may effect on the functionality of hematite material as a result of its insulating action [34]. By keeping this in mind, we have applied high heat treatment temperature during the processing stage to get rid of insulating maghemite phase and in order to have fundamental understanding of structural evolution which affects the device performance. On the other hand to get a detail picture on the variation of oxidation state and site symmetry of the material during thermal treatment, we have employed Fe K- edge XANES spectroscopy. Hereby, we are taking into account of the pre edge peak intensity ratio result for probing the oxidation state and site symmetry.

## II. Experiment

For the synthesis of iron oxide nano particles high purity (99%) iron (III) nitrate [$Fe(NO_3)_3.9H_2O$] and stearic acid [$C_{18}H_{36}O_2$ ] in the ratio of 1:2 were used as initial ingredients. The homogeneous solution of molten mixture was then heated at 125°C for 90 minutes which then formed a reddish brown viscous mass which was subsequently treated with tetrahydrofuran (THF). The powdery precipitates were collected through centrifugation and dried completely in a furnace supplied with ambient air at 70ºC. The dried precipitates were further subjected to heat treatment at 250ºC for holding time of 30 minute inside the electrically heating furnace to get the nano particles. The remainder of the precursor complex have been synthesized in a similar manner by annealing in systematic fashion from 300°C to 900°C to obtain the hematite phase. It was performed in non-isothermal manner by putting the precursor samples in $Al_2O_3$ crucibles in a Ceram-Aix furnace (FHT 175/30) using a heating rate of 300 K/h to the annealing temperature, a holding time of 30 min and a cooling rate of 300 K/h. Care was taken to slow down the cooling rate so as to allow the sample to heal oxygen vacancies, which are known to be introduced at high temperatures.

X-ray diffraction was carried out using $CuK_{\alpha}$ radiation; wavelength 1.54 Å. The phase composition was examined by analysing the powder X-ray diffraction pattern with PAN Analytical X'Pert PRO software. Transmission electron micrographs were obtained using a Philips CM30 Transmission electron microscope operated at 300 keV. The temperature of the sample stage was cooled with liquid nitrogen. Particle size distribution was determined by LS 230 particle size analyzer from Beckman Coulter by using the PIDS (Polarization Intensity Differential Scattering) signal of 40%. The lower resolution limit of this instrument is 40 nm. X-ray absorption spectra at the Fe K edge were collected at the Swiss-Norwegian beam line (SNBL) at the European Synchrotron Radiation Facility (ESRF), Grenoble, France. The electron energy in the storage ring was 6 GeV with a maximum current of 200mA. A Si(111) double-crystal monochromator was used for energy selection, which was detuned by 20–50% in order to suppress higher harmonic radiation. The intensities of the incident and transmitted X-rays were monitored with nitrogen-filled ionization chambers. The monochromator was scanned from 200 eV below to 800 eV above the Fe K-shell absorption edge (7111 eV). The Fe K-edge XAS were measured in the transmission mode at ambient temperature. A 7-µm thin Fe foil was used as a reference for X –Ray energy calibration. Nano particle powders were mixed with boron nitride in 1:5 ratio and then pressed to pellets for the X-ray absorption measurements. Background subtraction and normalization of the spectra was performed with standard procedures using the software program WinXAS [35].

## III. Results and Discussion

During the synthesis, magnetite in the dry state is oxidized to maghemite by air. It is known that ultrafine crystals of magnetite change over years from black to the brown of maghemite even at room temperature [27]. At temperature higher than 350ºC the transformation further proceeds to hematite. Also, the role of surfactants (also referred to here as ligands) such as stearic acid used in this case usually bind to the surface of the nanocrystals and give rise to a steric hindrance to aggregation [36]. This method has widely been used because of the ease and reproducibility of the synthesis, as well as the uniformity, high crystallinity, and monodispersity of the product. Also, because of the dependence of photocatalytic and photoelectrochemical properties on the nano particle size, significant efforts have been concentrated on the precise control of particle size distribution [37-38]. The capping group plays an important role in controlling the particle size.

The X-ray diffractograms of six samples heat treated at temperatures from 250°C to 500°C (Figure 1-A) showed the presence of both α and γ phases. On increasing heat treatment temperature, the as prepared powder sample at 250ºC undergoes complete decomposition of the organic compounds, and the hematite structure evolves along with phases of $\gamma$-$Fe_2O_3$. At 350ºC, the composition remains same and after that complete phase transformation of nano particles occurred. Above 350°C it is observed that α phase of the nano particles dominates the diffractogram. From this discussion it is evident that at higher heat treatment temperature $\gamma$-$Fe_2O_3$ completely transformed into $\alpha$-$Fe_2O_3$. This transformation of maghemite to hematite is considered as topotactic with the [111] and [110] axes of maghemite corresponding to the [001] and [110] axes of hematite [39]. The formation of hematite is favored at high temperatures [40]. It has been observed that the reflections coming from low temperature annealed samples were broadened suggesting the small size of the crystallite. Careful inspection of these reflections shows that they actually become broaden non-uniformly in case of samples from 250°C to 350°C. This non-uniform broadening was proposed originally to be due to shape anisotropy of the particles rather than to strains and faults broadening which remains in acicular particles after further heating at 600°C [41]. Initially the particles are smaller and irregular in shape as is evident from electron microscopy data. Upon heat treatment the larger, prismatic shaped nano particles formed in well distributed form at the expense of the smaller particles, likely by diffusion based growth similar to Ostwald ripening. As the temperature increased, the Bragg reflections systematically got sharper and it is significant from increase of the crystallite size along with decrease of full width with half maximum with heat treatment temperature (Figure 1-B). In the case of hematite derived from the thermal decomposition of akaganeite, a similar trend was observed [42]. This was due to the increase of crystallization with time and temperature plays a role in obtaining the solid phase. The increase in crystallite size can affect both stoichiometry and structure of the nano particle interior and surface or even can cause local or total phase transformation [43]. The phase transformation with increase in crystallite size generally occurs in order to obtain a stable phase by nanomizing the surface energy. From the energetic study of different set of iron oxide it was found that hematite is the most stable phase under aerobic surface condition in comparison to other phases whose enthalpy of formation is higher [44]. The surface energy is nothing but excess energy of the corresponding nano particles with respect to bulk material [45]. It is found to be very high in case of smaller nano particle with higher surface area. On the other hand, from the expanded view of the XRD profile (Figure1-B), it is found that the (104) and (110) Bragg reflections remain at the same position and shape was well maintained on increasing the temperature from 250°C to 500°C. At 300°C the peak got slightly shifted to lower Bragg angle possibly due to the repeated phase transformation of the precursor complex to maghemite and then to hematite during the nucleation and growth process of the nano particle formation. It might also be due to the presence of net lattice disorder as the particle sizes are quite small as evidence from the broad Bragg reflections.

However, after 300ºC the reflections mainly signify the hematite phase of the nano particle. In this case, the formation of the maghemite like structure is mainly restricted by the kinetic issue during the growth of hematite nano particle by the sintering process. This means that not only particle size but also the growth kinetics determines the structure of the nano particles [46]. The transformation of maghemite to hematite could also be attributed to the fact that with increasing particle size, maghemite ($\gamma$-$Fe_2O_3$) which is a defect spinel polymorph preferentially nucleates and hematite phase obtained on increasing the size. It is worth mentioning here that the spinel phase is thermodynamically less stable due to the higher surface energy of the nanocrystallites as the particle size got increased. The likely reason is that the spinel phase has a more open lattice structure, a lower number of broken bonds per unit surface, and lower surface polarity due to increased bond covalence [43].On the other hand, in case of nano particles heat treated from 550°C to 900ºC (Figure 2.A) showed the presence of $\alpha$ phase only. From the expanded view of the peak (110) and (104) as shown in Figure 2.B, we observe that the position and shape of the peaks remained identical up to 650ºC, but from 700°C up to 900 º C the (110) reflections split into two peaks. This peak splitting in case of 550°C to 900°C samples could be due to hexagonal distortion of the usual lattice structure of hematite or might be due to the imposing of compressive stress on the surface of nanostructure by the increasing oxygen content at high temperature. Besides these, at 900ºC both peaks finally changed its position to lower angle. This peak shift might be attributed due to formation of defects on the surface of hematite nano particle at high heat treatment. The signature of defect state in nanocrystalline material was well explained from the theoretical calculation of X-ray diffractogram results. From the calculated diffraction pattern, it was proposed that the nanocrystals displays peak shifts and broadening because of the presence of stacking fault as a defect in case of face–centered cubic crystalline geometry. In addition, it was also found that the location of stacking fault has largely influence the magnitude of the peak shifts [47]. In our study, we have observed such kind of stacking fault for 900 ºC heat treated sample as evident from the TEM image discussed in next section. At 900 ºC the nanocrystal structure was thought to be transformed from regular octahedral geometry to tetrahedral configuration which mostly signified the presence of magnetite phase at high temperature treated sample. This will be further confirmed with the quantitative analysis of the relative pre edge peak intensity ratio in the Fe – K edge XANES spectra.

In order to get visual information about the nano particles size, its crystallinity and morphology, high resolution transmission electron microscopy (HRTEM) was further applied. Representative images of four samples (250°C, 300°C, 600°C, and 900º C) are shown in Figure 3(A-F). We observe that the particle sizes range from 2-50 nm and show good crystalline character. Formation of rings in the selected area electron diffraction pattern (SAED) (Figure 3.A) of the sample annealed at 250°C signifies the polycrystalline nature. From the TEM images of 250°C and 300ºC sample (Figure 3.A, B) it was evident that particles were in disordered state whereas at high temperature (600°C) the particle shape became prismatic, and not round shape, which is standard morphology of small hematite crystals. Besides these from HRTEM image (Figure 3.D) we can see the well alignment of crystal plane with $d_{hkl}$ value of 2.521Å which corresponds to that of the lattice distance of the (110) plane. Finally in case of sample heat treated at 900ºC (Figure 3. E), particles preserved the prismatic shape and from the high resolution scale (Figure 3.F); we have observed that some step or terrace like pattern formed over the surface of the nano particle with rhombohedral shape which can be attributed due to the formation of stacking fault like defect A similar type of stacking fault has also been observed in case of SiC crystal grown by spontaneous nucleation sublimation method [48]

Figure 4.A shows the mean particle number size distribution of the nano particles synthesized at different annealing temperature. For the at 250ºC annealed samples, the width of the distribution was very broad in comparison to the samples annealed from 300°C to 900°C. This can be explained by taking into account of the formation of two different types of nano particles during the entire heat treatment step. The first type is the large prismatic shaped nano particles for high temperature annealed samples, and the second one is the small disordered particles with lower

annealing temperature. Most probably, that the large spherical particles are "mature" ones at the respective conditions of growth, and the small ones are some "immature seeds" from which the large "mature" particles grow. This type of growth process is usually called as diffused mediated growth or Ostwald ripening. This refers to a dissolution–precipitation based mechanism for particle growth (coarsening) [45]. Hereby small solid nano particles have surface atoms with elevated chemical potential with respect to larger nano particles because of the presence of excess energy associated with the surface. This increases the relative solubility of small particles, and larger particles grow preferentially. It has also to be noted that the seed particles are characterized by larger dispersion than mature particles. On average, the mean and median diameter systematically decreased by about 10% from 250°C to 900°C annealing temperature (Figure 4. A). In particular, no coarsening or growth of particles is observed. Instead, the nano particles actually shrink during heating. In contrast, visual inspection of the Bragg reflexes shows that their full width at half maximum decreases during heat treatment, revealing that the crystallites actually grow (Figure 4. B). We thus observe a growth of crystallites and a shrinking of primary nano particles. This phenomenon can be explained by the depletion of defects during temperature increasing. Further support for this will be shown by changes in the the pre-edges of the x-ray absorption spectra.

Figure 5 shows the normalized Fe-K edge XAS spectra of iron oxide nano particles heat treated at temperatures from 250°C to 900°C. These spectra can be divided into two regions, e.g., the pre-edge and post edge. The spectra are shown in three different energy resolution scales as in Fig. 5 (A-C). The pre-edge features in XANES spectra of transition metal K-edge are attributed to quadrupole transitions from 1s to 3d orbitals, dipole allowed transitions due to the 3d-4p mixing of the metal and 3d-p mixing between the metal atom and ligands through bonding and multiple scatterings involving the same atoms with different scattering paths [49]. Therefore, the positions of the pre-edge peaks directly reflect the crystal field splitting of 3d orbital sub-bands, and intensities of the pre-edge features are sensitive to the local oxygen coordination geometry of the metal atom. The pre-edge shifts with increasing oxidation state to higher energy. The number of transitions in the pre-edge depends on the oxidation state and coordination number [50]. In order to make a systematic and quantitative assessment of the changes in the pre-edge spectra, an accurate background subtraction is necessary, particularly since the pre-edge intensity is very small in comparison to the white line intensity. The adopted routine of smooth baseline interpolation appears reasonable, although one cannot entirely rule out a subjective component [51]. On systematic investigation of the pre-edge peak intensity, we observe that the intensity increases on increasing the annealing temperature. This is due to the change in oxidation state and site symmetry of the crystallite on respective sintered condition. The pre-edge peak intensity decreases with increasing coordination number. In the case of Fe, unlike Ti and Ni species, for example, the pre-edge energy does not shift as a function of coordination number [49].

The sample tempered at 250°C shows a broad pre-edge maximum at about 7113 eV and an intermediate plateau at around 7117 eV. Its relatively high pre-edge intensity suggests that $\gamma$-$Fe_2O_3$ is the dominant phase, in line with the XRD results. Similar strong pre-edge was observed by Chen et al. [52] in Fe oxide nano particles, who also observed that with increasing particle size the pre-edge features resembled more those of $\alpha$-$Fe_2O_3$. This is in line with the observations made on maghemite particles in general, i.e. that maghemite typically forms as small nano particles, whereas larger iron oxide particles with $Fe^{3+}$ transform to $\alpha$-$Fe_2O_3$ because of thermodynamical reasons, i.e. a nanomization of the free surface energy [53]. Usually the sample annealed from 500°C to 750°C showed Fe in the $Fe^{3+}$ oxidation state along with octahedral coordinated geometry for the respective structural pattern of hematite. In the 750°C sample, the structure was trigonal bipyramidal and again at 800°C the pre-peak intensity increased due to distortion of the local structure of the nano particle from trigonal bipyramidal to tetrahedral geometry. It is also worth mentioning that the usual trend of increase in intensity normally follows the order ferrihydrite > goethite > hematite [54]. This fact is quite obvious from the XRD pattern and might be due to the inverse correlation of the intensity of the pre-edge with the extent of centrosymmetry of the crystallographic site of Fe. Hence the most intense pre-edge peak was found for the sample annealed at $800°C$.

On deconvoluting the pre-edge peak of the respective XANES spectra, only two maxima are distinguishable in the pre-edge spectra which might be due to increasing site distortion. In the case of deconvoluted spectra as shown in Figure 6.A( for 250°C), two components obtained were mainly due to the distortion of the octahedron of oxygen atoms around $Fe^{3+}$, which is in agreement with theory. These two components were for high spin $Fe^{3+}$ ion with octahedral symmetry related to quadrupolar electronic transition to the $t_{2g}$ level. Besides these, pre-edge parameters are much more in agreement with those of the other $Fe^{3+}$ compounds including those for ferrihydrite for which high pre-edge intensity was erroneously thought to be related to the presence of tetrahedrally coordinated $Fe^{3+}$ [55]. The pre-edge intensity is more sensitive to site centrosymmetry with the most centrosymmetric Fe coordination having lowest intensity. For the Fe K edge, the pre-edge peak intensity is slightly higher in reduced samples for the same composition and there is an energy difference of about 0.5 eV between the edges of the first pre-edge peaks in the oxidized and reduced samples. The edge energy position (7113 eV of the first pre-edge peak for reduced samples is in agreement with that of $Fe_2O_3$ having all iron ions in the $Fe^{3+}$ oxidation state. The relative peak area of the normalized peaks (maxima) against different annealing temperature has also been studied and the results were shown in Figure 6.B. The normalized peak area due to first peak decreases linearly with increasing temperature and the area due to the second peak increases on increasing the sintering temperature. The shape of the pre-edge peak changes from a split to a normal singlet peak. This could be explained by considering the fact that this splitting of the pre-edge feature decreased with lower $Fe^{2+}$ coordination. That is with increase in heat treatment temperature both the oxidation state and symmetry of the nano particle changes from $Fe^{3+}$ octahedral to $Fe^{2+}$ tetrahedral environment. This means that at high heat treatment temperature the crystallite within the nano particle somehow displays faced centered cubic arrangement, which supports the formation of stacking fault defect and XRD peak shifting as discussed in XRD and TEM results. In case, the inversion symmetry of the transition metal is broken, which normally found for maghemite nano particle obtained from 250°C to 300°C as in this case, the pre-edge gains additional intensity due to the local 3d-4p wave function mixing, effectively allowing dipole transitions to the 4p character of the 3d-band. The pre-edge peak also increases slightly in energy with the valence state of the absorbing iron. The distinctive feature of the pre-edge peak that is used in this experiment is the fact that for octahedrally co-ordinated iron, the pre-edge peak is split into two components with a separation of 1.5 eV [56 -57]. This is caused by crystal field splitting of the ground state [58]. Peaks are associated with the $t_{2g}$ and $e_g$ transitions [59]. For tetrahedral coordination, the crystal field splitting between the $e_g$ and $t_{2g}$ levels is much less so they are not resolvable and appear as a single peak [56]. This was found to be the case even when only 60% of the iron ions were tetrahedrally coordinated and the rest were octahedrally coordinated. Detail of the pre-edge peak at the base of the edge shows that the oxides containing tetrahedrally coordinated iron, magnetite ($Fe_3O_4$), and maghemite ($\gamma$-$Fe_2O_3$) give a single peak whereas hematite ($\alpha$-$Fe_2O_3$) and the two oxyhydroxides, lepidocrocite ($\gamma$-FeOOH) and goethite ($\alpha$-FeOOH), all of which contain octahedrally coordinated iron, give a split peak.

## IV. Conclusion:

In summary, iron oxide particles in the size range of 5-50 nm have been synthesized by a non-aqueous soft chemistry route with different annealing temperature. The systematic analysis of the evolution in structural properties of nano particles synthesized at different heat treatment temperatures was studied by X-ray diffractometry. The XRD results show that at the lower annealing temperature nano particle comprise both the maghemite and hematite phase. As the temperature got increased up to 900°C the maghemite phase seemed to be converted to the hematite phase with the corresponding formation of stacking fault like defect at 900°C which validated the structural transition of hematite particle from regular hcp arrangment to fcc geometry. Also by detailed analysis of the (110) and (104) peak it was found that the peak position gets shifted and split upon increasing the corresponding synthesis temperature. This is primarily because of the local site distortion of the geometry of the nanocrystallite got distorted at the respective higher temperature. TEM results further showed that the samples synthesized at 250 $^{\circ}$C, 300 $^{\circ}$C, 600 $^{\circ}$C and 900$^{\circ}$C were all crystalline in nature with well developed lattice fringes as evident from high resolution imaging. The shape of the particles synthesized at low temperature was spherical whereas it becomes prismatic on increasing the heat treatment temperature. The selected area electron diffraction pattern of nano particles exhibited its polycrystalline nature. Finally the sample heat treated at 900$^{\circ}$C showed some step like pattern on the surface of the nano particle which is due to the formation of stacking fault like defect as evident from literature study. The particle size distribution study further signified that during the synthesis process two types of nano particle got formed and the distribution got sharper with increase in the heat treatment temperature which was due to the presence of mature particle at that temperature and it becomes quite broad for smaller seed particle from which growth of mature particle took place. Finally, from the pre-edge analysis of the Fe-K edge XANES spectra, it can be concluded that the pre-edge peak intensity increased with respect to the annealing temperature due to change in oxidation state and site symmetry of the nano particle. Quantitative analysis of the relative pre-edge peak area further signified the $Fe^{3+}$ octahedral to $Fe^{2+}$ tetrahedral transition in respective nano particle sample on increasing the sintering temperature.

**ACKNOWLEDGEMENT**

The research leading to these results received funding from the European Community's Sixth Framework Marie Curie International Reintegration Program grant n° 042095, Seventh Framework Program (FP7/2007-2013), Novel Materials for Energy Applications grant n°. 227179, Swiss NSF grants 200021-116688 (S.E.) and IZK0Z2-133944 (A.B.) and Swiss Federal Office of Energy contracts 153613-102809 and 153476-102691 (D.K.B.). These XANES experiments were performed on the Swiss Norwegian Beamline at the European Synchrotron Radiation Facility (ESRF), Grenoble, France.

**Figure Captions**

Figure 1: X-ray diffractogram of iron oxide nano particles obtained  by heat treatment of precursor complex  from   (A) 250-500 °C (B) expanded view of (104) and (110) peak in samples synthesized at  250 -500°C .

Figure 2: X-ray diffractogram of iron oxide nano particles obtained  by heat treatment of precursor complex  from  (A)550-900°C (B) expanded view of (104) and (110) peak in 550-900°C samples.

Figure 3: TEM images of samples synthesized at (A) 250°C (Inset: SAED pattern) (B) 300°C (C) 600°C and (E) 900°C. (D – F) The corresponding HRTEM images of 600 °C and 900°C samples.

Figure 4: (A) Nano particle size distribution measured by particle size analyzer (B) variation of mean and median parameter with synthesis temperature along with crystallite size calculated using the Scherer equation.

Figure 5: (A) Normalized Fe K-edge X-ray absorption spectra of $Fe_2O_3$ nano particles synthesized at different heat treatment temperatures. (B) XANES region of the normalized Fe K-edge X-ray

absorption spectra. (C) Pre-edge region of the normalized Fe K-edge X-ray absorption spectra.
(D) Deconvoluted pre – edge region of spectra from different set of nano particles by varying the temperature.

Figure 6: (A) Deconvoluted pre-edge peak of the respective XANES spectra for $250^{\circ}C$

(B) The relative peak area of the normalized peaks (maxima) against different annealing temperature.

Figure.1

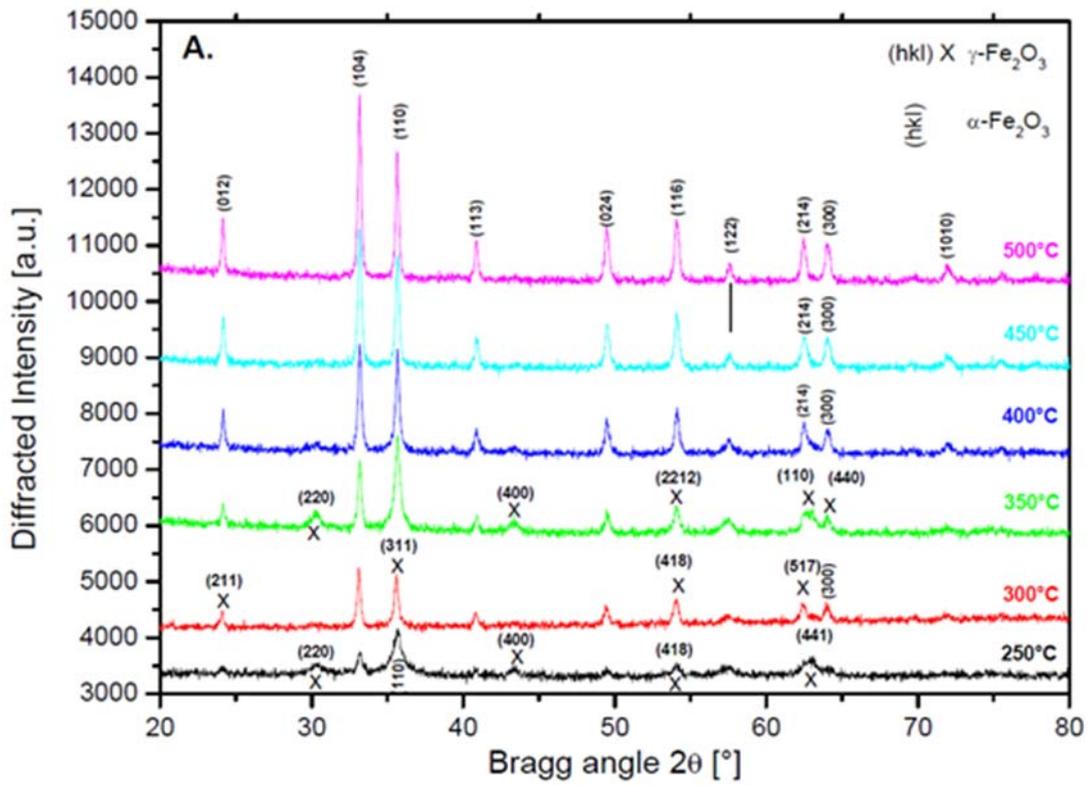

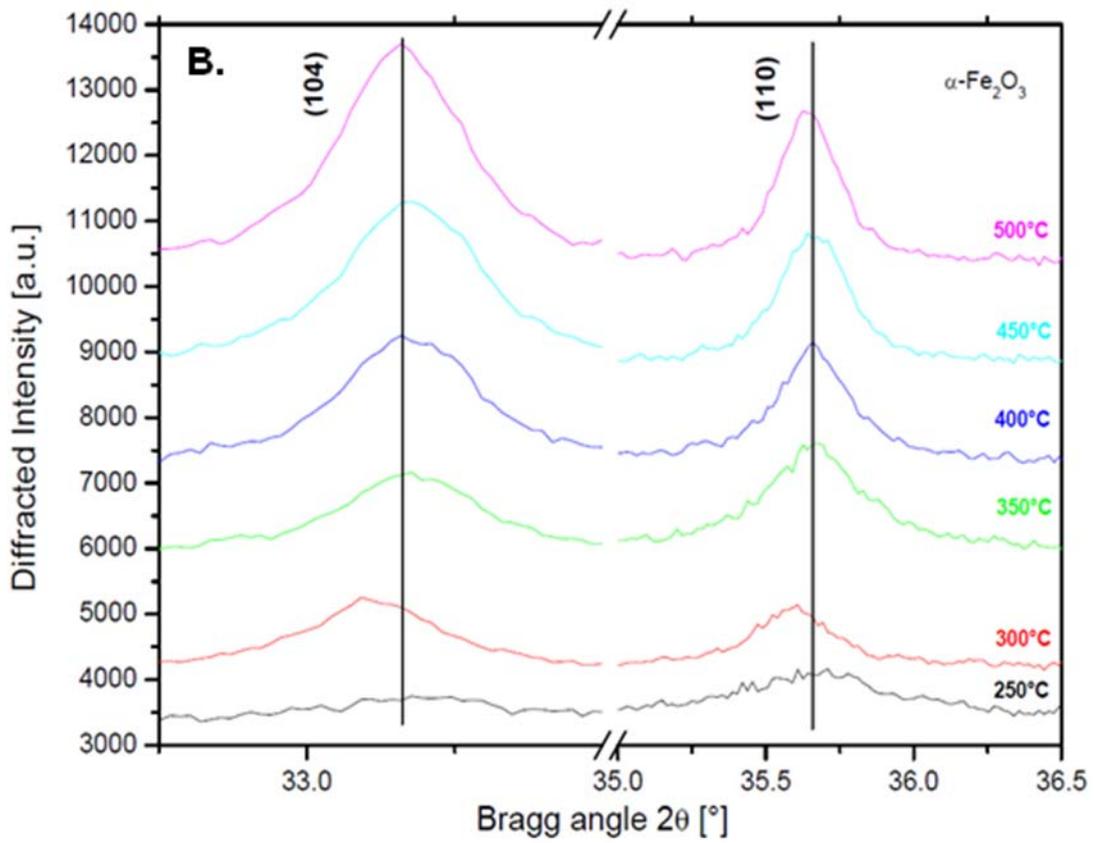

Figure 2

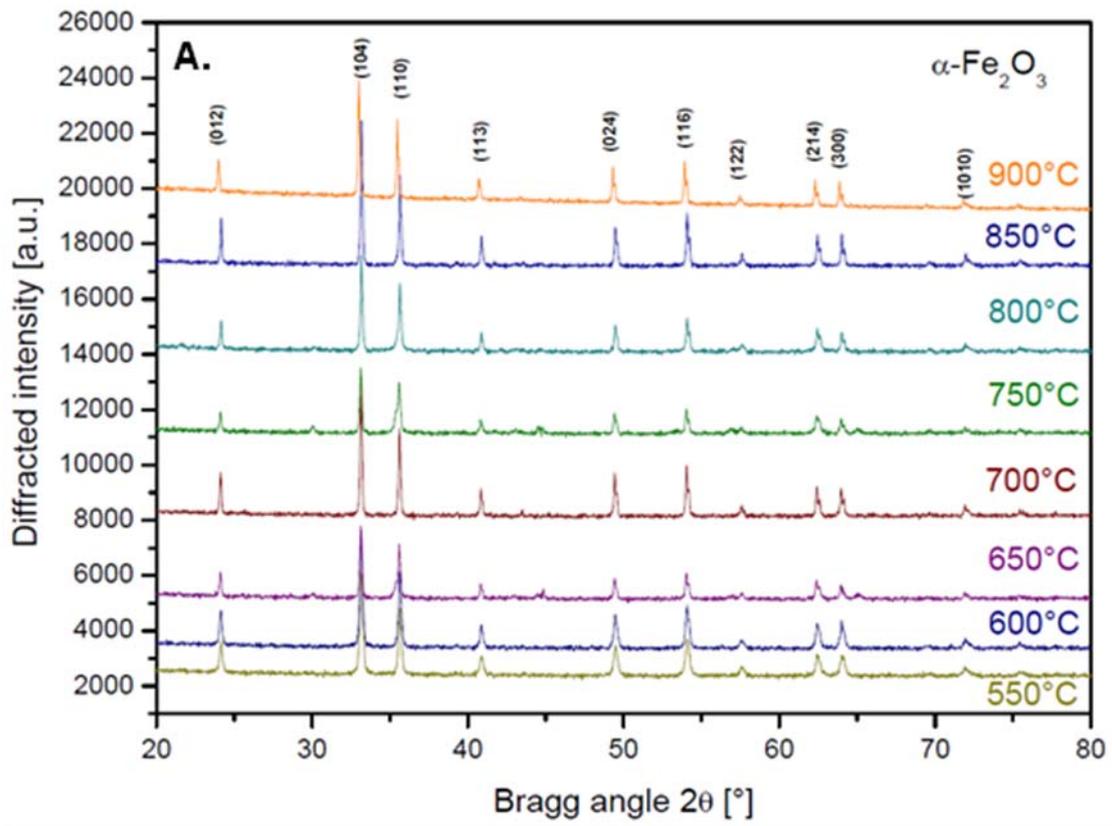

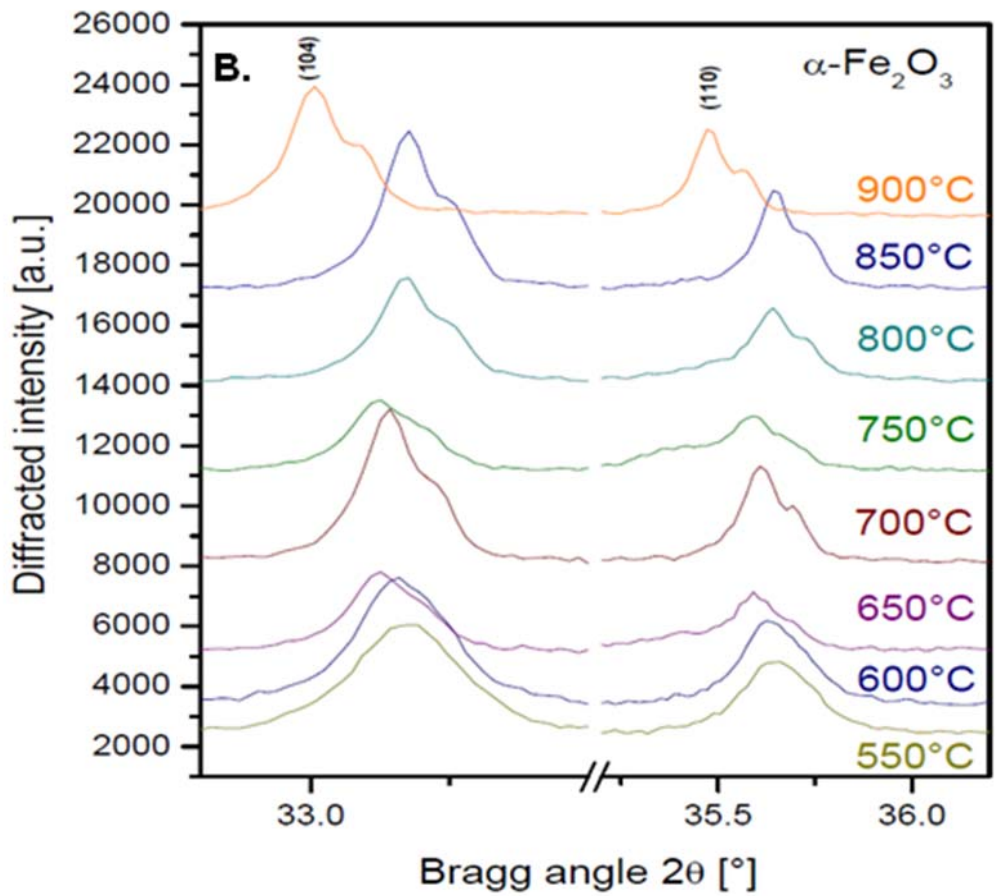

Figure 3.

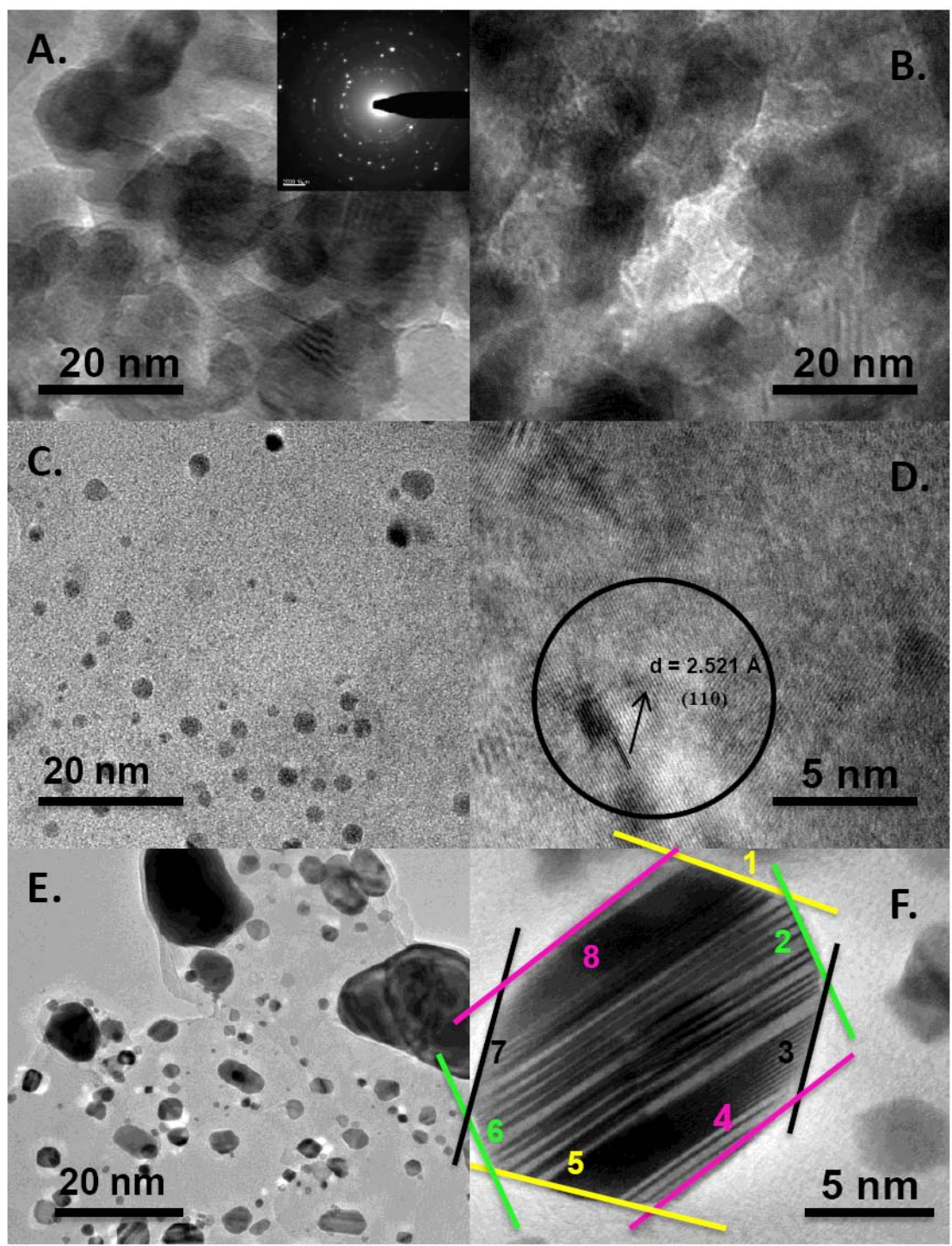

Figure 4.

A.

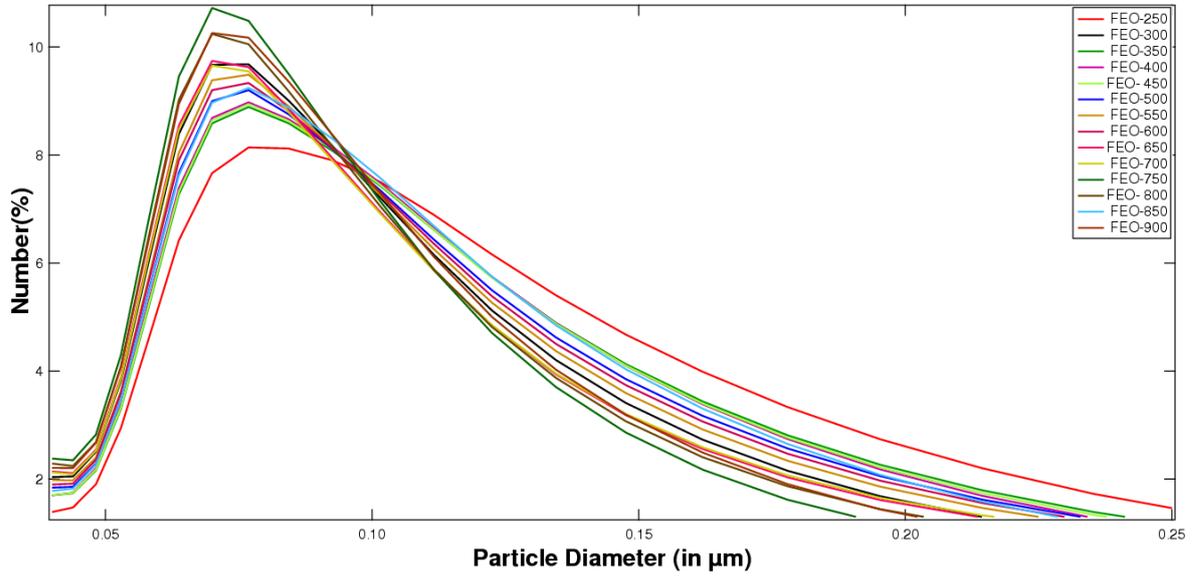

B.

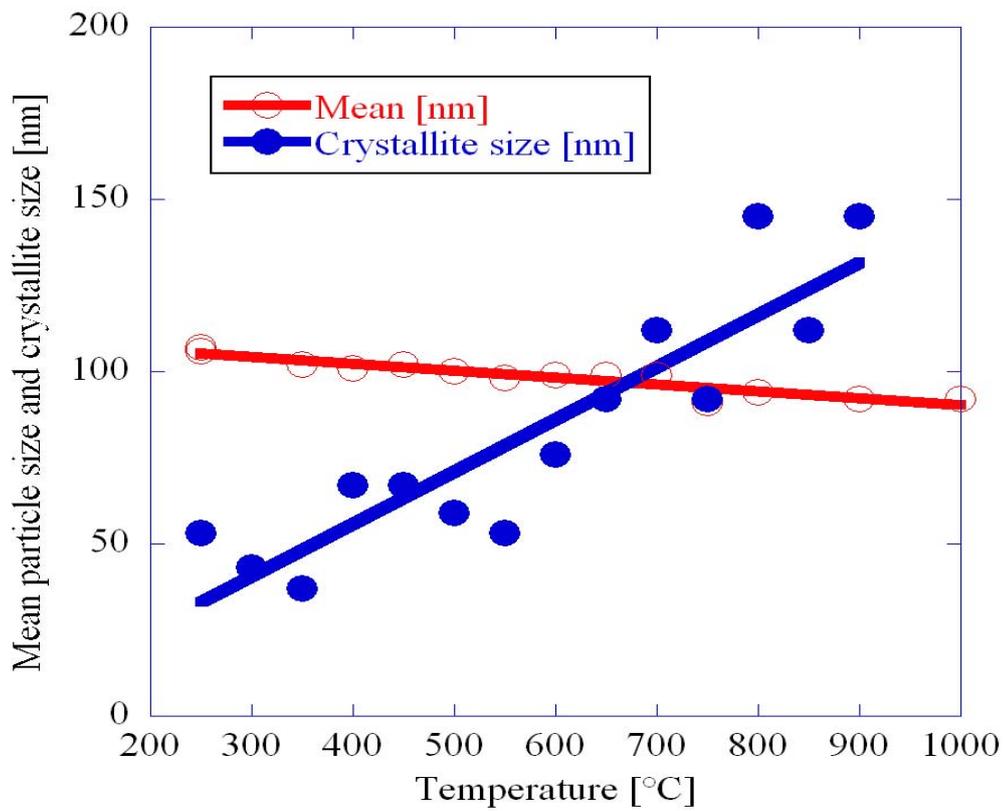

Figure 5

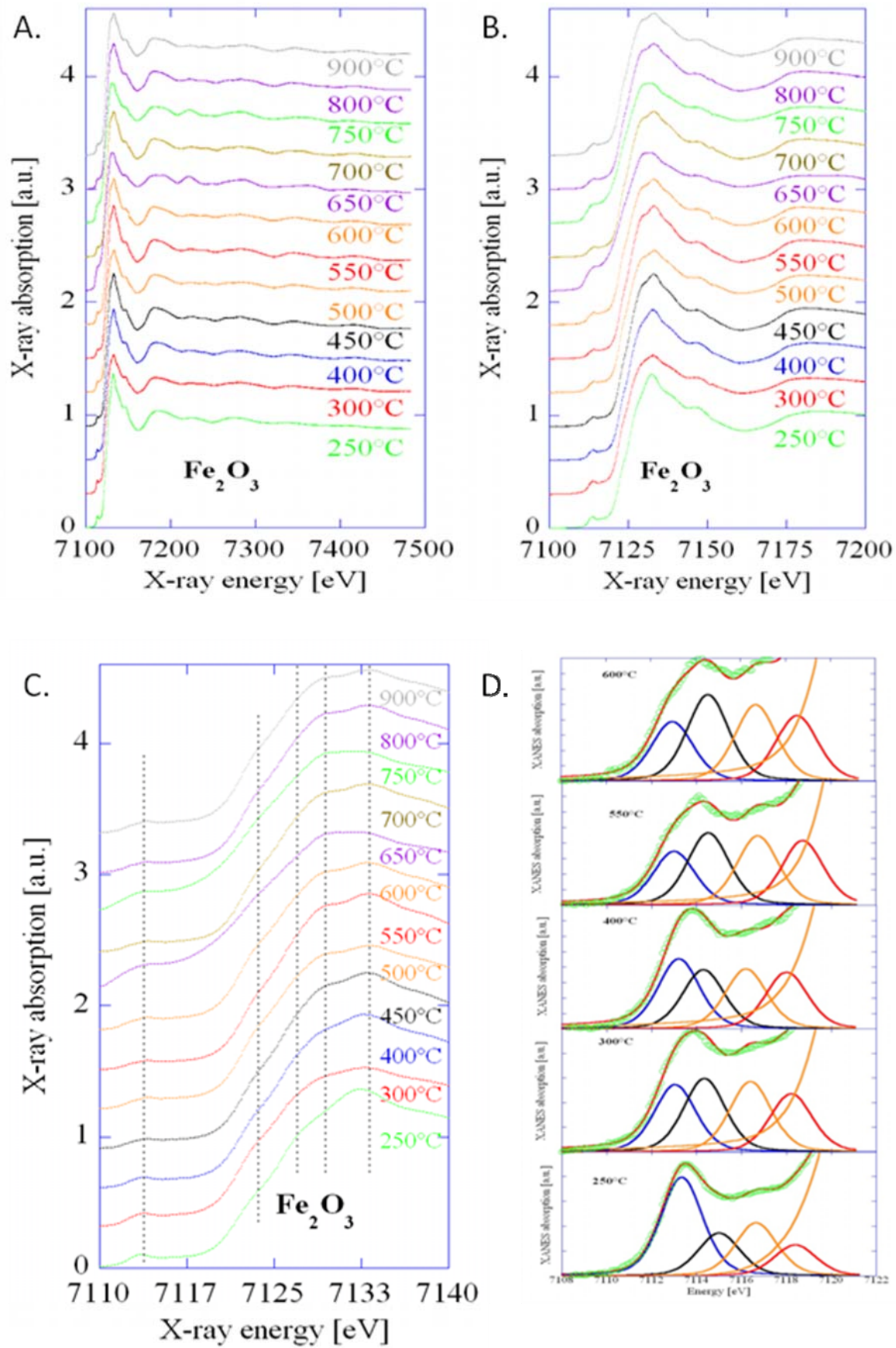

Figure 6

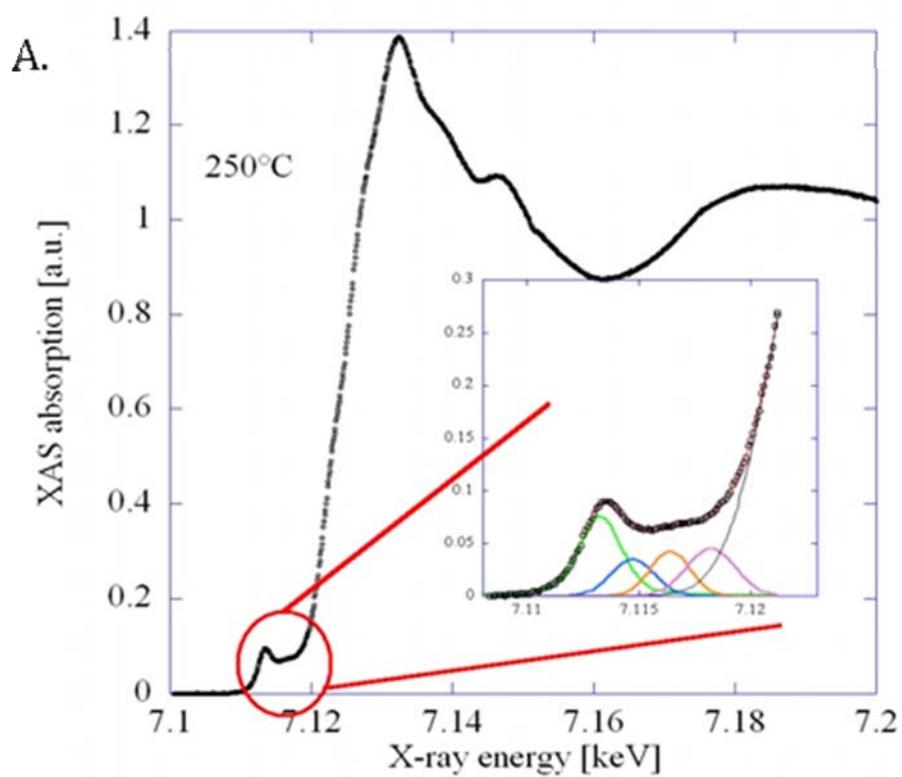

A.

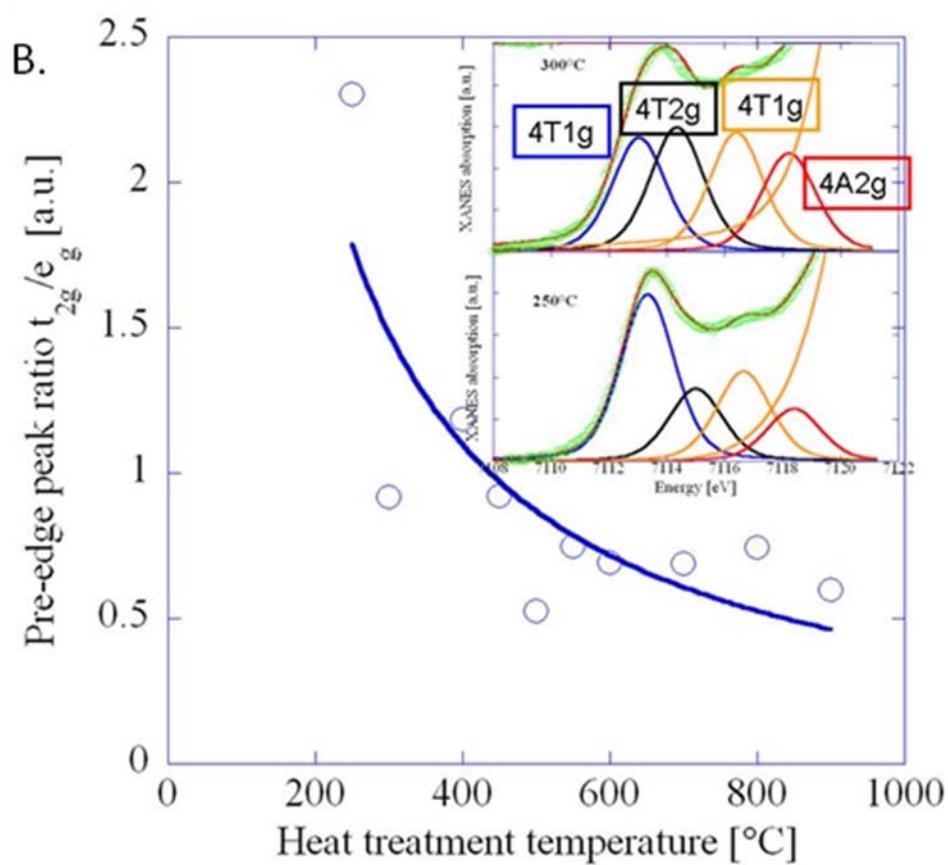

B.